\documentclass[useAMS,usenatbib]{mn2e}
\usepackage{graphicx,amsmath,multirow,amssymb}
\usepackage{subfig}
\usepackage{natbib}
\usepackage{txfonts}
\usepackage[usenames,dvipsnames]{color}
\newcommand{\comment}[1]{}

\def\simgt{\lower.5ex\hbox{$\; \buildrel > \over \F sim \;$}}
\def\simlt{\lower.5ex\hbox{$\; \buildrel < \over \sim \;$}}

\title[]{Dust from evolved stars: a pilot analysis of the AGB to PN transition}

\author[F. Dell'Agli et al.]{
F. Dell'Agli$^{1}$\thanks{E-mail: flavia.dellagli@inaf.it},
S. Tosi$^{1,2}$,
D. Kamath$^{1,3}$,
L. Stanghellini$^{4}$,
S. Bianchi$^{2}$,
P. Ventura$^{1,5}$,
\newauthor
E. Marini$^{1}$,
D. A. Garc{\'\i}a-Hern{\'a}ndez$^{6,7}$
\\
$^{1}$INAF, Osservatorio Astronomico di Roma, Via Frascati 33, 00077, Monte Porzio Catone, Italy \\
$^{2}$Dipartimento di Matematica e Fisica, Universit\'a degli Studi Roma Tre, via della Vasca Navale 84, 00100, Roma, Italy \\
$^{3}$Department of Physics and Astronomy, Macquarie University, Sydney, NSW 2109, Australia \\
$^{4}$NSF's NOIRLab, 950 Cherry Ave., Tucson, AZ 85719, USA\\
$^{5}$Istituto Nazionale di Fisica Nucleare, section of Perugia, Via A. Pascoli snc, 06123 Perugia, Italy\\
$^{6}$Instituto de Astrof\'isica de Canarias (IAC), E-38205 La Laguna, Tenerife, Spain\\
$^{7}$Departamento de Astrof\'isica, Universidad de La Laguna (ULL), E-38206 La Laguna, Tenerife, Spain
}

\begin{document}

\date{Accepted, Received; in original form }

\pagerange{\pageref{firstpage}--\pageref{lastpage}} \pubyear{2012}

\maketitle

\label{firstpage}

\begin{abstract}
We present a novel approach to address dust production by low- 
and intermediate-mass stars. We study the asymptotic giant branch (AGB) phase,
during which the formation of dust takes place, from the perspective of 
post-AGB and planetary nebula (PN) evolutionary stage. Using results from stellar evolution
and dust formation modelling, we interpret 
the spectral energy distribution of carbon-dust-rich sources currently evolving through 
different evolutionary phases, believed to descend from progenitors of similar 
mass and chemical composition. Comparing the results of different stages along the AGB to PNe transition, we can provide distinct insights on the amount of dust and gas released during the very late AGB phases. While the post-AGB traces the history of dust production back to the tip of the AGB phase, investigating the PNe is important to reconstruct the mass-loss process experienced after the last thermal pulse. The dust surrounding the post-AGB was formed soon after the tip of the AGB. The PNe dust-to-gas ratio is $\sim10^{-3}$, 2.5 times smaller than what expected for the same initial mass star during the last AGB interpulse, possibly suggesting that dust might be destroyed during the PN phase. Measuring the amount of dust present in the nebula can constrains the capacity of the dust to survive the central star heating. 

\end{abstract}




\section{Introduction}
AGB stars are efficient pollutants of the interstellar medium,
because their cool and dense winds prove a favourable environment for dust production. 
For example, these stars are known to significantly contribute to the chemical enrichment of galaxies as well as up to 40\% of interstellar medium dust \citep[e.g.][]{raffa14, goldman22}.

The modelling of dust formation by AGB stars has made significant advance due to the development of an interfaces
between stellar evolution modelling results and the chemo-dynamical description of the
winds \citep{ventura12, nanni13}, which allow for the calculation of the
dust production rate (DPR) during the entire AGB evolution. This approach
has been successfully applied to characterize the evolved stellar populations of
the Magellanic Clouds \citep{flavia14, flavia15, nanni19,
ester21} and Local Group galaxies \citep{flavia16, flavia19}.

Despite recent improvements, current dust formation models are still affected by several uncertainties regarding the 
description of the AGB phase. In particular, the treatment of the convective
borders and the mass-loss mechanism remain significant sources of uncertainty. 

The study of more evolved counterparts of AGB stars, such as post-AGB stars and PNe, makes a significant contribution to the determination of the dust budget from AGB stars. The majority of the dust released by AGB stars is produced during the final phases of their lifetime. As a result, this dust, or a significant portion of it, will remain in the vicinity of the central object during subsequent evolutionary phases. This dust is responsible for the infrared excesses observed in the spectral energy distribution (SED) of post-AGB stars \citep{devika15, devika22} and PNe \citep[e.g.,][]{letizia12}.

During the last decade, estimates of the mass and metallicity of the PN progenitors in the Magellanic Clouds (MC) and the Galaxy were estimated through the comparison of the chemical composition of the nebula with the prediction of the AGB stellar evolution models \citep{ventura15, ventura16, ventura17, stanghellini2022}. Recently, theoretical developments in modeling the transition from the AGB phase to subsequent phases \citep{marcelo16, devika23} have also enabled the characterisation of post-AGB sources \citep{devika23} based on physical parameters such as luminosity, effective temperature, and surface chemical composition. 

In this work, we make one step further providing a comprehensive interpretation of the observational information available from the final AGB stages to the PNe phase. We aim to demonstrate that the simultaneous study of stars in different evolutionary phases laying along the same evolutionary sequence (in the luminosity-T$_{\rm eff}$ plane) provides valuable insight into the amount of dust produced during the AGB phase, the cessation of dust production, and the transport of dust away from the star. This approach provides a global reading of the transition from the AGB, to the post-AGB, and PN phases, facilitating a deeper understanding of the production and evolution of dust during the late stages of a star's life.

\section{Different stars with the same progenitor mass}
The study of post-AGB stars and PNe offers a distinct advantage due to the reliability of luminosity-T$_{\rm eff}$ location of the evolutionary tracks on the HR diagram within these phases, unlike earlier evolutionary phases (i.e., AGB and prior to the AGB) that are subject to uncertainties related to mass loss and convection. 
The luminosity
of the stars remains approximately constant from the late AGB phases, following the last thermal pulse (TP), during the contraction phase through the
post-AGB phase, until the effective temperature reaches approximately 5$\times$\,10$^4$\,K. After that point, the luminosity decreases, as the white dwarf cooling sequence is approached.

We will capitalize on this for this study, therefore ensuring the simultaneous investigation of three sources from the Large Magellanic Cloud (LMC), specifically from the AGB, post-AGB and PNe evolutionary phases. As described in the following, from the data available in the literature it was possible to determine the effective temperature and luminosity for a wide number of evolved stars. These allowed us to carefully select three individual objects that, within the observational errors, are along the same evolutionary track; therefore, the chosen sources represent stars with the same initial mass.  Our objective is to
examine whether the DPR, a crucial parameter for assessing the dust production efficiency of AGB stars, is not only consistent with observations of AGB stars but also capable of explaining the observational data from post-AGB and PNe sources. By doing so, we aim to establish the compatibility of the AGB DPR across these different last stellar evolutionary stages. 

\begin{figure*}
\begin{minipage}{0.33\textwidth}
\resizebox{1.\hsize}{!}{\includegraphics{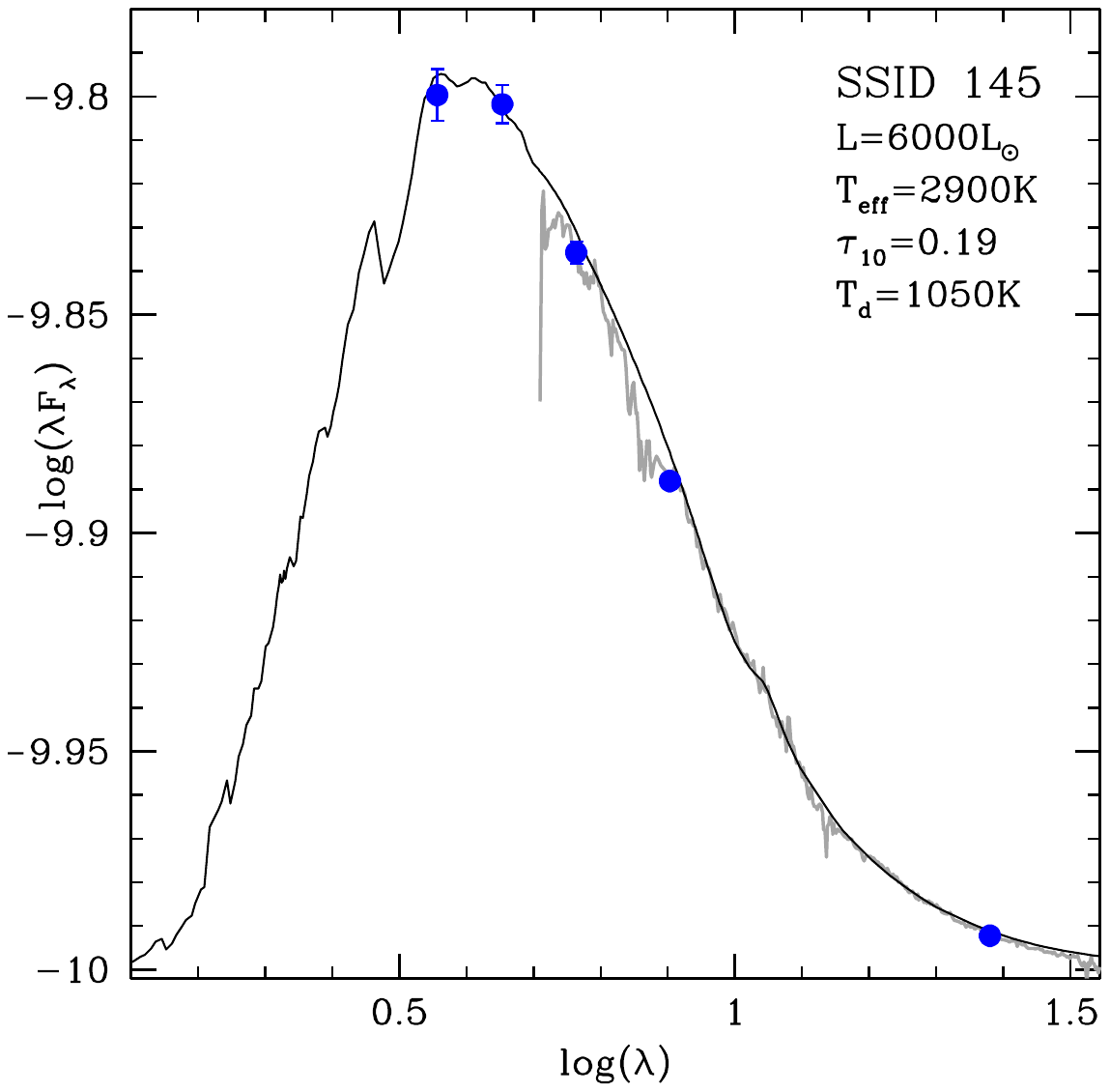}}
\end{minipage}
\begin{minipage}{0.33\textwidth}
\resizebox{1.\hsize}{!}{\includegraphics{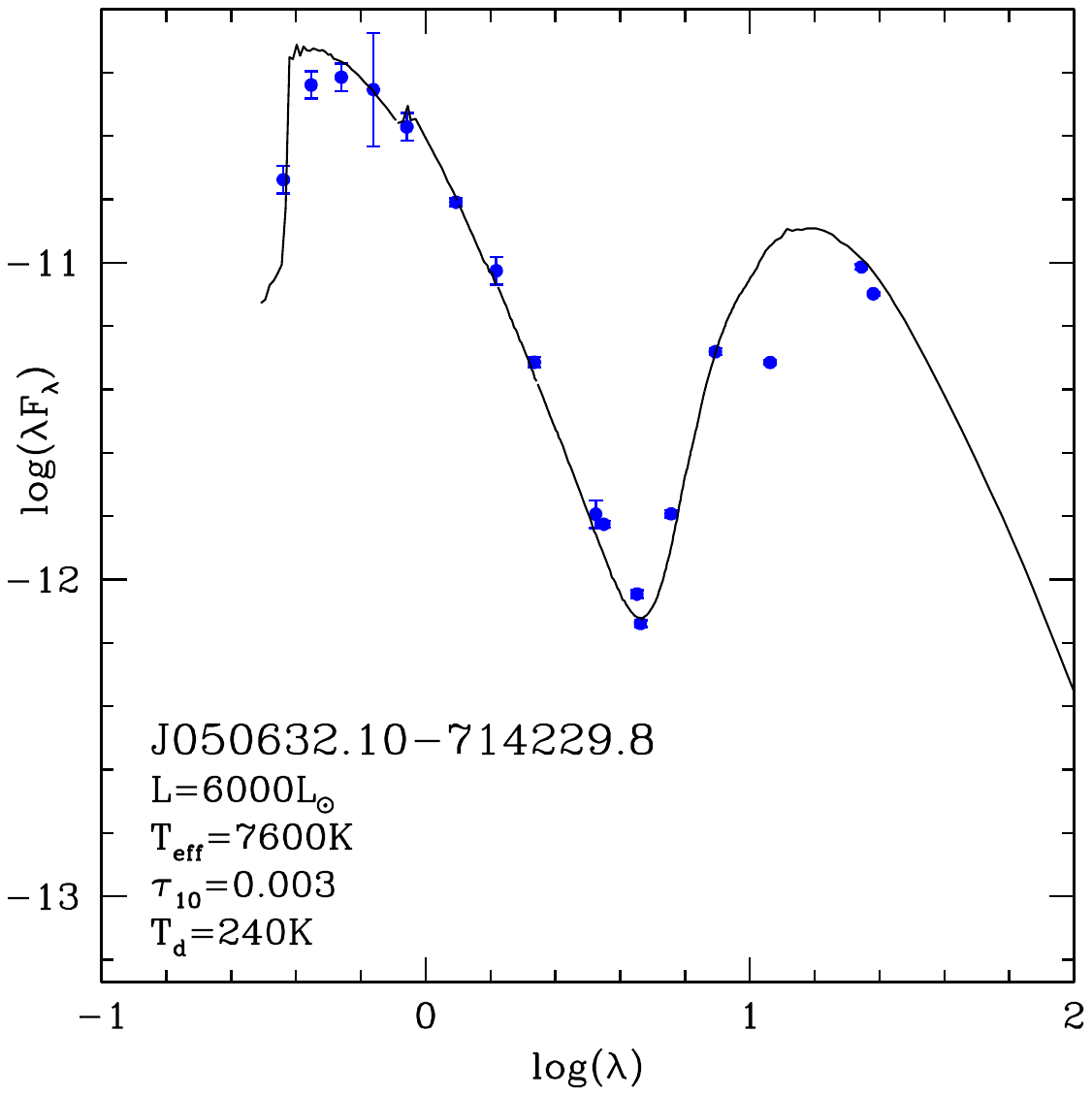}}
\end{minipage}
\begin{minipage}{0.33\textwidth}
\resizebox{1.\hsize}{!}{\includegraphics{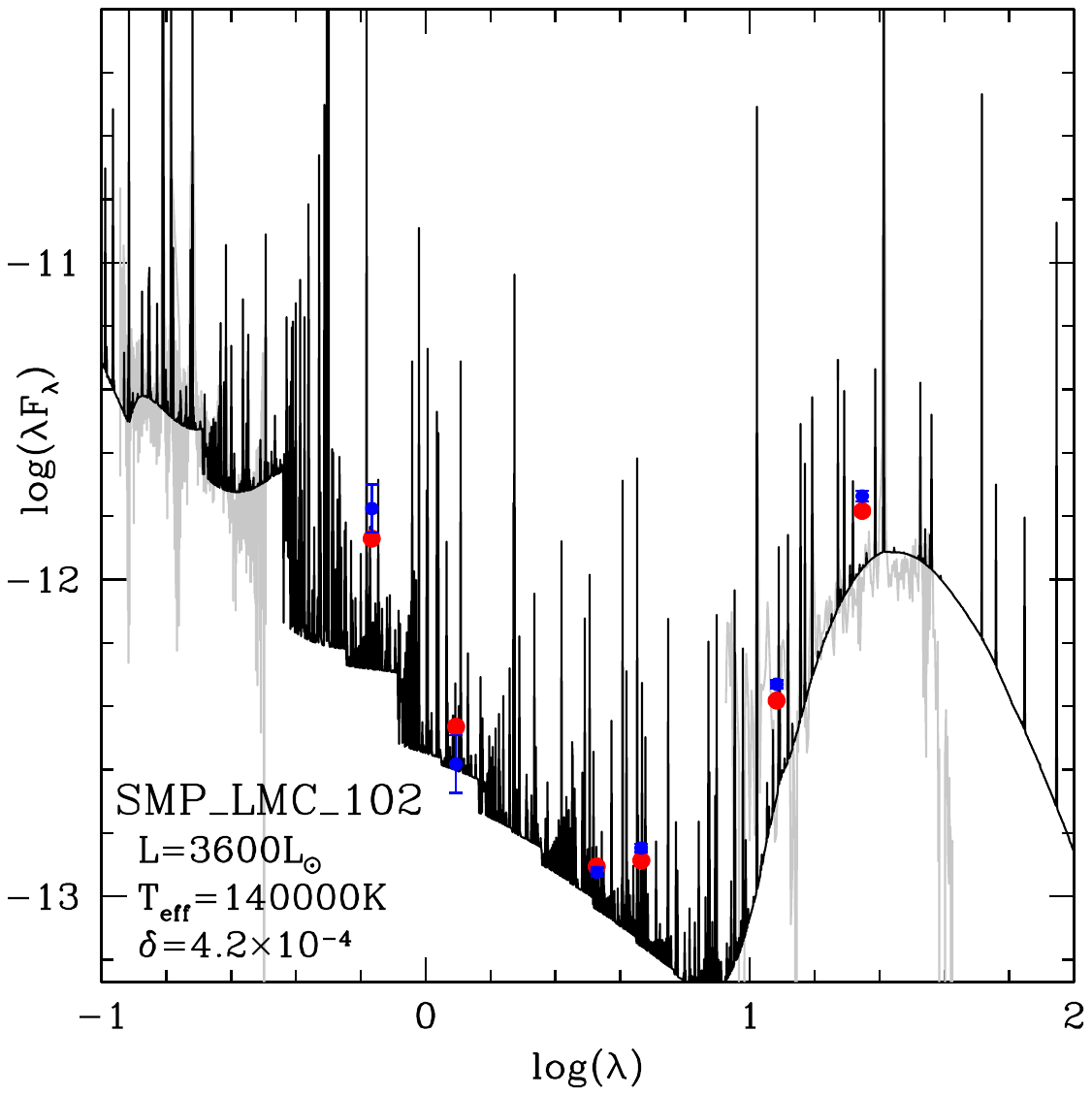}}
\end{minipage}
\vskip-40pt
\caption{Left: \textit{Spitzer}/IRS spectrum of SSID 145 (grey line), the results from \textit{Spitzer}/IRAC photometry 
(blue dots) and the best fit (black line) obtained by \citet{ester21}. 
Middle: photometric data of the source J050632.10-714229.8, taken from \citet{devika15},
are indicated with blue dots, overlapped to the best fit obtained by T22 (black line).
Right: the SED of SMP LMC 102, composed by the photometric data (blue dots) from
\citet{reid14, cutri12, lasker08}, the \textit{HST}/STIS UV spectrum taken from 
\citet{letizia05} and the \textit{Spitzer}/IRS spectrum from \citet{letizia07} (grey lines); the black line and red dots indicates the synthetic spectrum and photometry obtained in this work.}
\label{fall}
\end{figure*}

To constraints the AGB stars we consider the carbon star SSID 145 from the SAGE-Spec legacy survey of the LMC \citep{kemper10}. This sample of stars was investigated by
\citet{ester21}, who derived their luminosities and provided a characterisations 
in terms of mass and formation epoch of the progenitors. The \textit{Spitzer} spectrum (grey) and photometry (blue) of SSID 145 is shown in the left panel of Fig. \ref{fall}. 

Regarding post-AGB stars, we focus on the source J050632.10-714229.8 within the sample identified 
by \citet{devika15}. These post-AGB stars were thoroughly investigated 
by Tosi et al. (2022, hereafter T22), who derived the luminosity of the individual 
sources and explored the AGB to post-AGB transition, as well as the dynamics of gas and dust. The SED of this source is reported in the central panel of Fig. \ref{fall}.

\begin{figure*}
\begin{minipage}{0.48\textwidth}
\resizebox{1.\hsize}{!}{\includegraphics{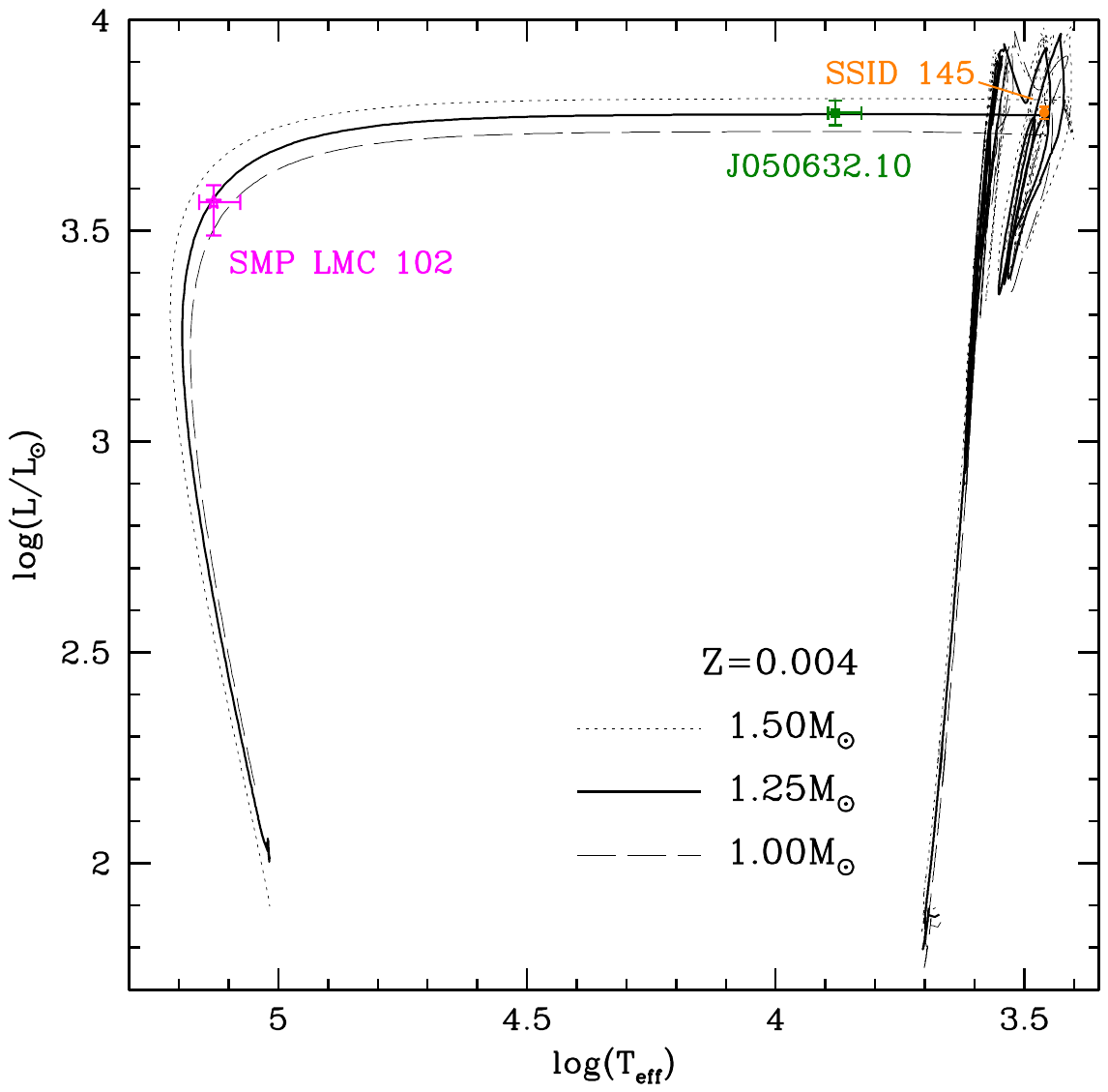}}
\end{minipage}
\begin{minipage}{0.48\textwidth}
\resizebox{1.\hsize}{!}{\includegraphics{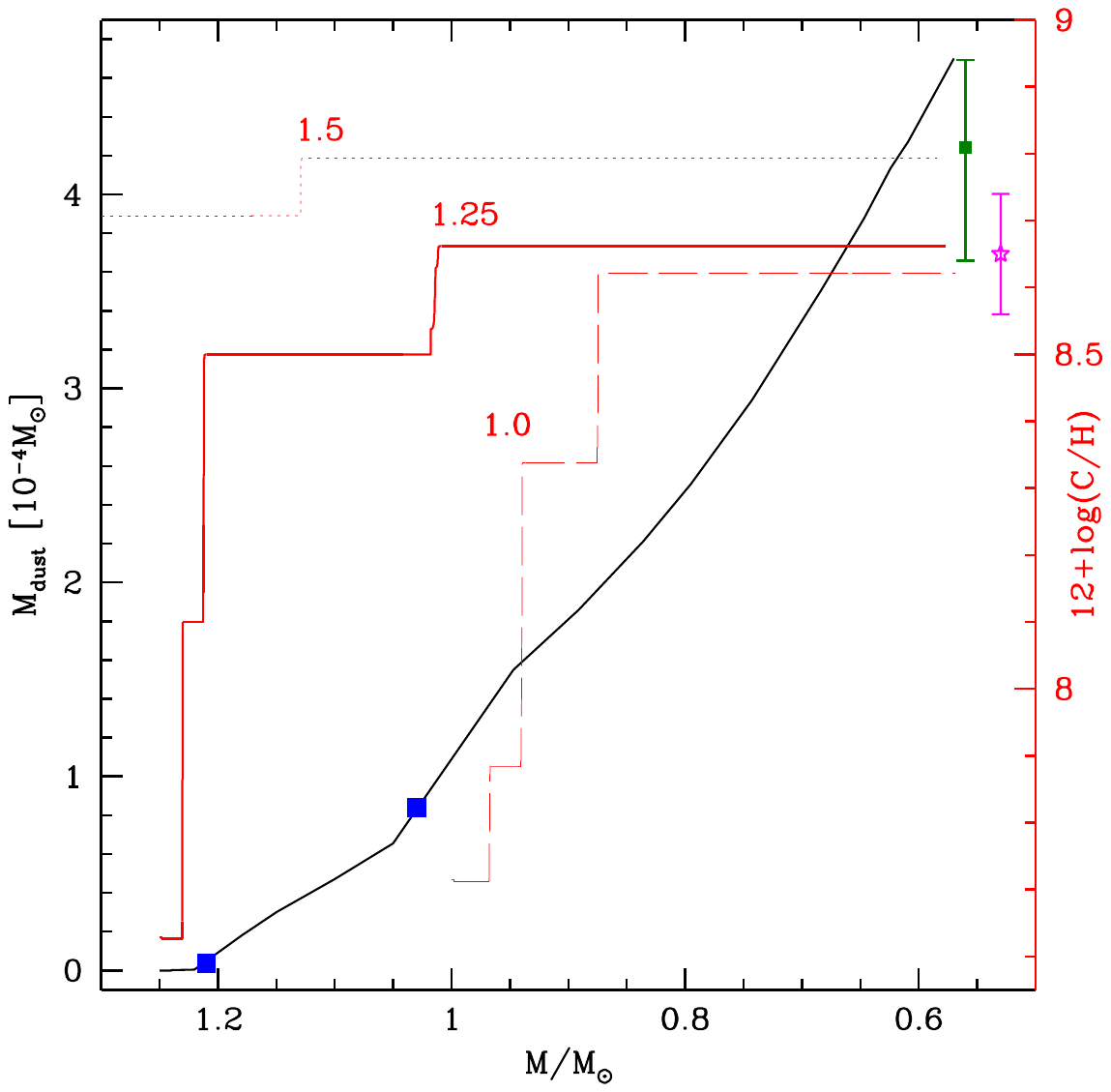}}
\end{minipage}
\vskip-60pt
\caption{Left: Evolutionary tracks on the HR diagram of the 1.0, 1.25 and $1.5~{\rm M}_{\odot}$ model
stars of metallicity $Z=4\times 10^{-3}$; the location of the AGB (orange full dot), post-AGB (green full square) and PN (magenta open star) sources are reported according to the results discussed in section \ref{snapshot}. Right: AGB variation of the dust mass released by the $1.25~{\rm M}_{\odot}$ model star (black line, scale on the left) and of the surface carbon abundance (red, 
scale on the right expressed in 12+log(C/H)) for the models considered in the left panel, as a function of the current mass of the star. Blue squares 
indicate the last two TPs, while the dots represent the 12+log(C/H) abundances of the post-AGB star J050632.10-714229.8 \citep[green full square;][]{vanaarle13} and the PN SMP LMC 102 \citep[magenta open star;][]{letizia05}.} 
\label{f125}
\end{figure*}

For the PNe, we consider the spherical and elliptical sources in the sample presented by \citet{letizia07}, that were further 
characterized by \citet{ventura15} on the basis of the chemical abundance measurements. In the present analysis we consider all the photometric and 
spectroscopic data available to reconstruct the SED from the UV to the mid-IR. This allow us to derive the properties of the central object including the thermal and chemical stratification of the gas cloud, as well as the dust density, following the methodology described in the following section. This initial analysis led us to identify the source SMP LMC 102, which, when plotted on the HR diagram, lies on the same evolutionary track as the post-AGB and the AGB counterparts mentioned earlier (see Fig. \ref{f125}). 
For this source we considered the photometric data available from the literature \citep{reid14, cutri12, lasker08}, the \textit{HST}/Space Telescope Imaging Spectrograph (STIS) \citep{letizia05} and the \textit{Spitzer}/Infrared Spectrograph (IRS) \citep{letizia07} spectra, as shown in the right panel of Fig.\ref{fall}. The \textit{HST}/STIS spectrum and the photometry have been corrected both for Galactic foreground and for the LMC extinction applying the same law and coefficients described in section 2.5 of \citet{letizia05}.

In the following section we present
the methodology used to determine the physical parameters characterizing  
each of the three sources, while the results are summarized in section \ref{snapshot}.

\section{Physical and numerical input description}
\label{input}

To determine the 
progenitor mass compatible with the physical parameters of  
the three selected sources, we calculated evolutionary sequences from the pre-main sequence 
to the AGB and post-AGB phases, extending until the PN stage. For this, we used 
the ATON stellar evolutionary code  \citep{ventura98}, incorporating the numerical updates
discussed in \citet{devika23}. These updates were specifically designed to accurately describe the 
transition to the post-AGB and PNe stages, ensuring reliability of our results.
We also modelled dust formation in the stellar winds throughout the entire AGB phase and
during the initial stages of the post-AGB evolution, following  
the schematisation described in \citet{fg06}. We summarize the main features of the model adopted and how it is coupled to the ATON code for stellar evolution in the Appendix, while all the relevant equations and details can be found in \citet{ventura12}.
This procedure allows for the calculation of a set of evolutionary tracks which describe the evolution of the central and the dust production process for different intial mass models and evolutionary stages. 

Using the DUSTY code \citep{nenkova99}, we computed the synthetic SEDs able to best reproduce the observed photometric and spectroscopic data of the individual AGB and post-AGB stars selected for this study, as described in \citet{ester21} and \citet{tosi22}. The set of parameters obtained characterizes each source during the specific evolutionary stage observed, independently of the modeling. We then compared the main parameters with our set of evolutionary tracks, which are discrete in mass and evolutionary time steps. The luminosity and effective temperature constrain the progenitor mass and the evolutionary stage, while the optical depth (at $\lambda=10~\mu$m) $\tau_{10}$, the dust composition, and dust temperature characterize the circumstellar dust. The Spitzer spectra available for the AGB sources allow for a comprehensive determination of the dust composition, including factors such as the fraction of SiC as opposed to amorphous carbon, graphite, and/or iron dust components (see Table 2 in \citet{ester21}). Conversely, the photometric data available for the post-AGB stars serve primarily to differentiate between carbonaceous dust and silicates. The optical depth is primarily influenced by the abundance of carbon/silicate dust within the circumstellar envelope, with minor species such as SiC, alumina, or iron dust playing a secondary role. Consequently, the comparison between the optical depth derived from SED fitting and the values predicted by the models provides indication on the amount of carbon/silicate dust that has formed.

To build the synthetic PN SED for the comparison with the observational data points,
we used the spectral synthesis code CLOUDY \citep{cloudy}. The computation is performed assuming a black body as a central source and a spherically symmetric nebula. The following inputs are taken form the literature: the electron density, N$_{\rm e}=1000$\,cm$^{-3}$, as determined by \citet{letizia05}, the effective temperature T$_{\rm eff}=131000 \pm 12400$\,K, from \citet{eva03}, and the inner radius of the photoionization zone, log(R$_{\rm in}$)=17.72\,cm of the nebula\footnote{We considered the radius R$_{phot}$ determined by \citet{shaw01} on the basis of the OIII emission to be able to set R$_{\rm in}$ rather than leaves it as a free parameters. Considering that the photoionization zone starts in an inner region of the nebula than the OIII emission, we scaled R$_{phot}$ by a factor 0.7 to obtain the R$_{\rm in}$, as shown e.g. Figure 2.6 in Osterbrock \& Ferland, 2006.}. The hydrogen density is assumed to be constant throughout the nebula. We adopted the solar abundances by \citet{lodders03} to be consistent with the ones used to compute the evolutionary tracks and we scaled it according to the overall metallicity estimated by \citet{ventura15}. For the dust grains we assume the presence of amorphous carbon with the refractive index as given by \cite{rouleau91} and the size distribution based on the work of \cite{mathis77}. The luminosity of the central star (CS), the thickness ($\delta$R) of the nebula and the amount of dust (expressed as number of carbon grains per unit hydrogen) are our free parameters to be determined. We run CLOUDY repeatedly in order to obtain the best agreement with the observational data, i.e. the photometry and the UV and IR spectra, as shown in the right panel of Fig. \ref{fall}. To have a consistent result we compare the intensity of the UV lines and the electron temperature as computed by CLOUDY with the results obtained by \citet{letizia05}. With this procedure we can constrain the luminosity of the CS and estimate the mass of gas and dust-gas-ratio present within the nebula. To estimate the uncertainties, we vary each parameter, keeping the others fixed, within a range of values able to produce synthetic photometry (red points in Fig. \ref{fall}) that deviates less than 5\% from the observational errorbars of the photometric data.

\begin{table*}
	\centering
    \renewcommand{\arraystretch}{1}
    \setlength{\tabcolsep}{6pt}	
    \caption{Top: parameters determined from the comparison between the observational data and the synthetic SEDs. Bottom: stellar evolution model parameters.}
	\label{table1}
	\begin{tabular}{ll|ccccc} 
\hline
\hline
\multicolumn{1}{l}{SED} &
\multicolumn{1}{l}{} &
\multicolumn{1}{l}{} &
\multicolumn{1}{l}{} &
\multicolumn{1}{l}{} &
\multicolumn{1}{l}{} &
\multicolumn{1}{l}{} \\
\multicolumn{1}{l}{} &
\multicolumn{1}{l}{Source} &
\multicolumn{1}{c}{Phase} &
\multicolumn{1}{c}{L / L$_{\odot}$} &
\multicolumn{1}{c}{$\tau_{10}$} &
\multicolumn{1}{c}{$\delta$} &
\multicolumn{1}{c}{M$_{\rm gas}$ / M$_{\odot}$} \\
\hline
& SSID 145 & AGB & 6000$\pm$150 & $0.19\pm0.02$ & -- & --\\
& J050632.10-714229.8 & post-AGB & $6000\pm 400$ & $0.003\pm0.0004$ & -- & -- \\
& SMP LMC 102 & PN & $3600^{+1000}_{-800}$ & -- & $(4.2\pm1)\times 10^{-4}$ & $0.37\pm0.04^{(a)}$ \\
	\hline
    \hline
\multicolumn{1}{l}{Stellar evolution model} &
\multicolumn{1}{l}{} &
\multicolumn{1}{l}{} &
\multicolumn{1}{l}{} &
\multicolumn{1}{l}{} &
\multicolumn{1}{l}{} &
\multicolumn{1}{l}{} \\
\multicolumn{1}{l}{} &
\multicolumn{1}{l}{M$_{\rm  i}$ / M$_{\odot}$} &
\multicolumn{1}{c}{Z} &
\multicolumn{1}{l}{$\dot{\rm M}$\,\,[M$_{\odot}$yr$^{-1}$]} &
\multicolumn{1}{c}{$\tau_{10}^{\rm TAGB}$} &
\multicolumn{1}{c}{$\delta_{\rm AGB}^{\rm last\,TP}$} &
\multicolumn{1}{c}{M$_{\rm gas}^{\rm last\,TP}$ / M$_{\odot}$} \\
\hline
& 1.25 & 0.004 & 1.7$\times 10^{-5}$ & 0.34 & 10$^{-3}$ & 0.4 \\
	\hline
    \hline
    \multicolumn{7}{|p{\linewidth}|}{Note: $\delta$ is the dust-to-gas ratio; $^{(a)}$ is the total gas mass of the nebula.}
	\end{tabular}
\end{table*}

\section{Snapshots along the transition}
\label{snapshot}
In this section we present for each of the three sources the results from the comparison between the observational data and the synthetic SEDs, from which the main physical and chemical parameters were obtained.
These parameters are summarized in Table \ref{table1}, while in Fig.~\ref{fall} we present the observational data set for the three sources considered, along with the results from the synthetic SED. We then compare these parameters with the ones obtained from the ATON code for stellar evolution and dust formation, whose some of the evolutionary tracks are shown in the left panel of Fig.~\ref{f125}.

The left panel of Fig.~\ref{fall} shows the SED of SSID 145 and the synthetic spectrum from which
\citet{ester21} derived a luminosity ${\rm L} = 6000\pm 150~{\rm L}_{\odot}$, 
$\tau_{10}=0.19\pm0.02$, and concluded that the dust around the star 
is made up of amorphous carbon, with traces of silicon carbide and MgS. 
Comparing these values with the ones obtained from the evolutionary models, this source was interpreted as the progeny of a 
$1.25~{\rm M}_{\odot}$ star, currently evolving through the final AGB phases (see Fig. 5 in \citet{ester21}. Looking at the age-metallicity relation of the LMC \citep[e.g.][]{harris09}, the typical metallicity of stars formed 1-1.5Gyr ago is Z$=1-4\times10^{-3}$.

The middle panel of Fig.~\ref{fall} shows the synthetic SED and the observational data of the post-AGB
source J050632.10-714229.8. From this comparison, T22 derived 
${\rm L} = 6000\pm 400~{\rm L}_{\odot}$ and $\tau_{10}^{pAGB}=0.003\pm0.0004$, due to the presence of 
amorphous carbon dust located at a distance $2.74\times 10^5~{\rm R}_{\odot}$ from 
the CS. Based on the comparison with the evolutionary tracks, T22 concluded that J050632.10-714229.8 descends from a 
$1.25^{+0.2}_{-0.15}~{\rm M}_{\odot}$ progenitor. The measurement [Fe/H]=-0.4 \citep{devika15} is compatible with a metallicity $Z\sim4\times 10^{-3}$. In the right panel of Fig.~\ref{f125}, we show the comparison between the 
carbon abundance of J050632.10-714229.8 given in \citet{vanaarle13} (green square) and the variation of the surface carbon abundance during the final phases of the AGB evolution for three model stars of mass 1.0, 1.25 and 1.5$~{\rm M}_{\odot}$ (red lines).

In the right panel of Fig.~\ref{fall}, we show the synthetic SED of the PN SMP LMC 102, obtained as explained in section \ref{input}. The derived luminosity were found to be $3600^{+1000}_{-800}~{\rm L}_{\odot}$. To reduce this uncertainty, we compared the measured abundance of 12+log(C/H) with the prediction from the evolutionary tracks, as shown in the right panel of Fig.~\ref{f125}. The surface carbon abundance of stars with masses $\ge 1.5~{\rm M}_{\odot}$ is too high to be compatible with the PN SMP LMC 102 carbon abundance; we can also exclude all the stars with masses $< 1$\,M$_{\odot}$ that do not become carbon stars. We therefore estimate a CS mass of $1.25^{+0.15}_{-0.25}~{\rm M}_{\odot}$. This results is rather robust, as stars in this mass range are not expected to be exposed to strong mass loss during the RGB, and also they do not experience any HBB, whose description is sensitive to convection modelling. The values of the core mass and luminosity obtained are within the values found by \citet{marcelo16} for stars of same mass and metallicity above (Z=0.01) and below (Z=0.001) the value used here. Furthermore, there is a good agreement among different evolutionary codes for the final carbon abundance predicted in the low-mass regime. This has been extensively investigated by \citet{ventura18} at solar metallicity, but we also tested it in the metal-poor regime. We considered Z=0.001 case for which several models are available, even if slightly lower than the metallicity analysed in this work. Considering a 1.25-1.3${\rm M}_{\odot}$ the FRUITY \citep{sergio11} and MONASH \citep{fishlock14} predictions for the final carbon surface abundance are within $\pm0.1$ dex from the ATON value, 12+log(C/H)= 8.65. 

Regarding the nebula, the thickness $\delta$R and the mass of the gas cloud are estimated
to be $0.06\pm0.003$ pc, and $0.37\pm0.04~{\rm M}_{\odot}$, respectively. 
The comparison between the synthetic SED and the observational data indicates that the IR excess is due to the presence of carbonaceous dust with a 
grain density of $\rm{n}_{\rm d}=(2.3\pm0.5)\cdot 10^{-11}\rm{n_{H}}$ and a gas-to-dust 
ratio $\delta_{\rm PN}=(4.2\pm1)\times 10^{-4}$. The chemical abundances of this source, already discussed, by \citet{ventura15} are compatible with a metallicity $Z\sim4\times 10^{-3}$.

\section{The AGB, post-AGB, PN timeline}

The AGB and post-AGB sources discussed in the previous section have the 
same luminosity within the observational uncertainties, which, based on the evolutionary sequences computed as described in section \ref{input}, suggests a progenitor star with an initial mass of $1.25~{\rm M}_{\odot}$ and metallicity $Z=4\times 10^{-3}$. 

The evolutionary tracks of 1.0, 1.25 and 1.5$~{\rm M}_{\odot}$ model stars with this metallicity are shown in the
left panel of Fig.~\ref{f125}, while on the right panel we show the AGB evolution of the
surface carbon for  (in red) and of the mass of dust released (in black).
Results from the stellar evolution modelling of a $1.25~{\rm M}_{\odot}$ model indicate that the age
of this star is around 3.5 Gyr, and that the AGB phase, lasting 
$\sim 1.3$ Myr, is characterized by the occurrence of 7 TPs, after which  
the evolution to the post-AGB phase begins.
The last three TPs are followed by third dredge up episodes, which favour the gradual increase
in the surface carbon, until reaching $[$C$/$Fe$] \sim 1$,
following which the star enters the post-AGB evolution.

During the carbon star stage, the accumulation of carbon in the surface regions of the star and the
simultaneous expansion and cooling of the external layers favour the formation of
significant quantities of dust, mostly composed by amorphous carbon grains,
which grows until reaching sizes of the order of $0.15~\mu$m. The IR emission in the SED increases, and the optical
depth grows, until $\tau_{10} \sim 0.35$ is reached at the tip of the AGB (TAGB). 
The overall amount of dust produced during the AGB
phase is $4.7\times 10^{-4}~{\rm M}_{\odot}$. If we restrict the attention
to the late phases following the last TP, the dust mass produced is 
$4\times 10^{-4}~{\rm M}_{\odot}$, while $\sim 0.45~{\rm M}_{\odot}$ of gas 
is ejected into the interstellar medium: these quantities correspond to an
average dust-to-gas ratio $\delta_{\rm AGB}^{\rm last\,TP} \sim 10^{-3}$.

During the evolutionary phases from the last TP until the TAGB,
$\tau_{10}$ increases from $\sim 0.05$ to $\sim 0.35$, while the luminosity
of the star is $6000~{\rm L}_{\odot}$. These values are consistent with
those derived in section \ref{snapshot} for SSID 145, which we therefore
believe is currently evolving through the final AGB phases, towards the TAGB.

J050632.10-714229.8 represents the post-AGB counterpart of SSID 145. T22 deduced
that the current IR excess observed is related to a
region around the star populated by carbonaceous dust, located at a distance of 
$\rm R_{dust}=2.74\times 10^5~{\rm R}_{\odot}$ from the CS, characterized by an optical depth 
$\tau_{10}=3\times 10^{-3}$. We considered the hypothesis that the IR excess present in the SED of the post-AGB is due to the last dust produced and released at the end of the AGB phase or immediately after, before the temperature increases and prevents the condensation. The methodology introduced by T22 (see the Appendix for more details) allows to estimate the \textit{onset} time when the dust was formed based on the comparison with the evolutionary tracks and dust formation model.

At the TAGB the $1.25~{\rm M}_{\odot}$ model star is characterized by effective 
temperature ${\rm T_{eff}} =2800$ K, stellar radius $320~{\rm R}_{\odot}$, 
mass-loss rate  
$\dot{\rm M}=1.7\times 10^{-5}~{{\rm M_{\odot}}}/$yr, $\tau_{10}=0.34$.
We ant to identify the evolutionary stage after the start of the contraction towards
the post-AGB phase when the dust currently observed was released. We will refer to it as the \textit{onset}, making the hypothesis that it is when the effective temperature increased to 3000 K and the
stellar radius dropped to R$_{\ast}=280~{\rm R}_{\odot}$. At the $onset$ time 
we find that $\dot{\rm M}=1.4\times 10^{-5}~{\rm M_{\odot}}/$yr and that the dust is
mainly located at a distance R$_{onset}\sim 10~{\rm R}_{\ast}$ from the
centre of the star, with $\tau_{10}^{onset}=0.24$. If we apply the scaling relation of 
the optical depth with distance proposed by T22, we find the current 
optical depth $\tau_{10}=\tau_{10}^{onset}\times \rm R_{\textit{onset}}/\rm R_{dust}=0.24\times 10~{\rm R}_{\ast}/(2.74\times 10^5~{\rm R}_{\odot}) \sim 3\times 10^{-3}$. This value is
in agreement with the quantity derived from the analysis of the IR excess 
of the SED (section \ref{snapshot} and Table 1). We can assume that, evolving at constant luminosity, the time needed by the dust to reach the distance R$_{dust}$ correspond to time required by the effective temperature to increase from 3000 K to the current estimated
value of 7600 K is $2500$ yr; if we assume an average velocity of the gas+dust
wind of 5 km$/$s \citep[see e.g.][]{he14}, we find that the dust should be currently located at a
distance $\sim 2.6\times 10^5~{\rm R}_{\odot}$ from the star, once more in
agreement with the results summarized in section \ref{snapshot}. The derived velocity is lower than those estimated for C-rich AGBs by \citet{martin2002}: this is consistent with the fact that the aforementioned \textit{onset} time is after the start of the contraction to the post-AGB, when dust production rate is lower, and so the expansion velocity.

According to the interpretation proposed here, SMP LMC 102 represents the 
PN counterpart of SSID 145 and J050632.10-714229.8.
The mass of the gas cloud surrounding the CS is $\sim 0.4~{\rm M}_{\odot}$,
which is equivalent to the overall gas lost by the star after experiencing the
last TP: we therefore propose that only this gas is currently and still trapped in the nebula,
whereas that released before the last TP was lost. The IR excess, clearly visible in the SED 
(see right panel of Fig.~\ref{fall}), is related to the presence of significant
quantities of carbonaceous dust, consistent with the analysis of SSID 145 and 
J050632.10-714229.8. The estimated dust-to-gas ratio given in section \ref{snapshot}
was $\delta_{\rm PN}=4.2\times 10^{-4}$. This value is 2.5 times smaller than the dust-to-gas ratio computed for the last interpulse of the AGB phase, $\delta_{\rm AGB} \sim 10^{-3}$. It could suggest that a fraction of the dust released during the final AGB phases is destroyed during the PN evolution \citep[see e.g.][]{vanhoof2000}. Further investigations on a wider sample of PNe should be needed in order to test this hypothesis.

\section{Discussion}
The results presented here highlight the advantages of simultaneously studying sources that are likely to be on a similar stellar evolutionary track but are in 
different evolutionary phases, ranging from the late AGB to the PN stage. 
With the availability of models that incorporate dust formation during the AGB phase, this approach emerges as a promising and valuable tool for assessing the expected dust budget from low and intermediate mass stars. 

Owing to uncertainties in the description of the TPs, this method is applicable only in the final AGB stages after the
last TP, when the luminosity attains the values characterizing the following 
post-AGB evolution. However, this limitation 
is not a major issue when considering dust production, since most 
of the dust is produced during the late inter-pulse phases. This is especially true in the low-mass domain, where significant amounts
of dust form only after the last TP, when the star becomes a C-star. 

In the present work we found that the study of post-AGB stars and PNe, when coupled with the modelling
of the evolution and dust formation of the AGB phase, can provide valuable insights to this kind of investigation, from two different, 
complementary perspectives. The analysis of the SED of post-AGB stars 
allows to reconstruct the properties of the stars at the tip of 
the AGB, when the physical conditions are most favourable to the condensation
of gas molecules into solid grains: this is important to determine the 
largest DPRs experienced by the stars across the 
AGB phase. While the study of post-AGB sources provides detailed information regarding the very late AGB phases only, the investigation of PNe gives a more
complete picture of the mass-loss and dust formation process that occurred 
during the final part of the AGB evolution. Interpreting the SED of PNe enables the  estimation of the amount of gas 
and dust trapped within the nebula surrounding the star. Comparing these 
results with those obtained from stellar evolution and dust formation modelling during the
AGB phase facilitates the understanding of: a) how much of the gas released by the star
during the late AGB phases is still trapped in the nebula; and b) the
percentage of dust that survived to sublimation.

In the specific example discussed here, we find that all the
gas released by the star after the occurrence of the last TP is currently 
trapped in the nebula, while only $\sim 40\%$ of the dust produced during 
those phases is still present in the nebula. This result could indicate that part of the dust has been destroyed during the heating process of the central star. However, further investigations are needed to prove this hypothesis and to determine whether this behaviour is general, or if other factors may alter this scenario.
For example, it would be interesting to verify if in carbon stars of higher mass, where the dust formation 
process is more efficient, significant fractions 
of dust may be lost via sublimation. This would necessitate a downward
revision of the dust budget estimated for these stars. On the other hand, 
the stronger radiation pressure associated with the increased dust production 
might facilitate the escape of the gas+dust wind from the AGB phase, resulting in smaller amounts of gas present in the nebula and allowing higher fractions 
of dust to survive. To address all these questions, it is crucial to apply the proposed approach to a wider sample of sources covering a larger range of initial masses. Also, it is interesting to see the results of a similar approach applied to stars at different initial metallicity than the LMC sequence, for example by using SMC of Galactic stellar sequences such as the one presented here. Future observations with the James Webb Space Telescope would further help in widening and enriching the variety and accuracy of the IR observational sample.

\section*{Data availability}
The data underlying this article will be shared on reasonable request to the corresponding author.

\section*{Acknowledgements}
F.D.A. and P.V. acknowledge the support received from the PRIN INAF 2019
grant ObFu 1.05.01.85.14 (“Building up the halo: chemo-dynamical tagging in
the age of large surveys”, PI. S. Lucatello) and from the INAF-GTOGrant 2022 entitled "Understanding the formation of globular clusters
with their multiple stellar generations" (P.I. A. Marino). F.D.A and M.E. are supported by the INAF-Mini-GRANTS 2022 ("Disclosing the stellar dust production in the Milky Way", PI. E. Marini). DK acknowledges the ARC Centre of Excellence for All Sky Astrophysics in 3 Dimensions (ASTRO 3D), through project CE170100013. DAGH acknowledges support from the State Research Agency (AEI) of the Spanish Ministry of Science and Innovation (MICINN) under grant PID2020-115758GB-I00. This article is based upon work from European Cooperation in Science and Technology (COST) Action NanoSpace, CA21126, supported by COST. This publication makes use of data products from the Wide-field Infrared Survey Explorer, which is a joint project of the University of California, Los Angeles, and the Jet Propulsion Laboratory/California Institute of Technology, funded by the National Aeronautics and Space Administration.







\appendix

\section{Model for dust production and wind dynamic}
In this section we make a summary of the model adopted to describe the dust formation and expansion of the dust during the transition from the AGB to the PN phase. Firstly we briefly describe the dust formation model adopted to compute the amount of dust produced in the circumstellar envelope of evolved stars during the AGB phase. The approach couple the description by \citet{fg06} for dust formation in stellar winds with the ATON stellar evolution models, as described in details in \citet{ventura12}. We assumed that the wind expands isotropically from the surface of the star with a constant velocity, described on the basis of the mass and momentum conservation. As the wind expands, the temperature cools down and the gas particles enter the condensation region. We follow the growth of the dust particles on the basis of the gas density and the thermal velocity in the inner border of the condensation region. When a certain amount of dust is formed, the extinction coefficient in the equation of momentum conservation increases and the radiation pressure on the dust particles accelerates the wind. Mass, luminosity, effective temperature, mass-loss rates and the surface chemical composition are the physical and chemical parameters, calculated by the stellar evolution code at several stages of the AGB phases, that enters in the dust formation computation. The surface chemical composition, resulting form the detailed description of nuclear and convective processes occuring during the AGB phase, determines the main dust species formed in the wind. In oxygen-rich environments (C/O$<$1) silicates are the main dust species formed and alumina dust (Al$_2$O$_3$) in a smaller quantity; in carbon-rich circumstellar envelopes (C/O$>$1) carbon dust is the major dust compound, with a minor quantity of silicon carbide possibly formed. The grain sizes, the density and type of dust formed determines the optical depth $\tau_{10}$ (see Table \ref{table1}), physical parameters that enters in the radiative transfer computation needed for the comparison with the observational data. 

In order to characterize the post-AGB source J050632.10-714229.8 we prosecuted the computation of the ATON evolutionary tracks after the TAGB, as presented by \citet{devika23} and \citet{tosi22}. To determine the formation history of the dust present in the surroundings of this object we applied the methodology presented by \citet{tosi22} and summarized here. We considered the hypothesis that the IR excess present in the SED of the post-AGB is due to the last dust produced and released at the end of the AGB phase or immediately after, before the temperature increases and prevents the condensation. We checked the consistency between the results obtained from dust formation modelling and from the SED (see Table \ref{table1}) considering the scaling relation introduced by \citet{tosi22}. We computed the optical depth which would have characterised the star if it was observed after the TAGB. To check compatibility with the current optical depth derived from SED fitting, we applied the following scaling relation introduced in T22:
$\tau_{10}=\tau_{10}^{onset}\times R_{dust}/R_{onset}$. In this formula, R$_{onset}$ is the distance of the dusty region from the centre of the star at the time when the dust was produced (typically five
to ten stellar radii from the centre of the star), while R$_{dust}$ is the
current distance of the dust from the post-AGB star. We evaluated the reliability of the hypothesis that the dust was released at a given evolutionary phase by comparing $\tau_{10}$ with the values derived from the SED fitting, reported in Table 1. We can work on this hypothesis estimating the time scale needed by the dust to move to the current location. Since the luminosity is constant from the TAGB to the post-AGB phase, we can think only on the basis of the timescale needed to the star to reach the effective temperature measured in the post-AGB star. We derived the velocity with which the wind would move since the dust was released until
the present epoch by dividing the derived $R_{in}$ by the time interval required for the effective temperature of the star to increase from the value at the considered evolutionary phase until the current value. The latter time was calculated based on stellar evolution modelling.
\label{lastpage}

\bsp	
\end{document}